\documentstyle[aps,twocolumn]{revtex}

\input psfig
\begin{document}
\title{Angle Dependence of the Transverse Thermal Conductivity in YBa$_2$Cu$_3$O$_7$ single crystals:
Doppler Effect vs. Andreev scattering} 
\author{R. Oca\~{n}a and P. Esquinazi} 
\address{Abteilung  Supraleitung und
Magnetismus, Institut f\"ur Experimentelle Physik II, 
Universit\"at Leipzig, Linn{\'e}str. 5, D-04103 Leipzig,
Germany}  
\maketitle 

\begin{abstract}

We have measured the transverse thermal conductivity $\kappa_{xy}$ of twinned and untwinned 
YBa$_2$Cu$_3$O$_7$ single crystals as a function of angle $\theta$ between the 
magnetic field applied parallel to the CuO$_2$ planes and the heat current direction, 
at different magnetic fields and temperatures. For both crystals we 
observed a clear twofold variation in the field-angle dependence of  $\kappa_{xy}(\theta) = 
- \kappa^0_{xy}(T,B) \sin(2\theta)$.
We have found that the oscillation amplitude $\kappa^0_{xy}$ depends on temperature
and magnetic field. Our results show that 
$\kappa^0_{xy} = a B \ln(1/(bB))$ with the temperature- and sample-dependent 
parameters $a$ and $b$. We discuss our results in terms of Andreev scattering of quasiparticles 
by vortices and a recently proposed theory based
on the Doppler shift in the quasiparticle spectrum.   
\end{abstract}

To the main effects produced by the $d$-wave symmetry of the superconducting order
parameter one includes the influence of  the extended quasiparticle (QP) states - associated with a gapless
structure (nodes) - on the mixed state of the superconductor in a magnetic field \cite{vol,kop}. 
In a semiclassical way, the effect of a magnetic field 
can be taken into account by introducing a Doppler shift (DS) in the 
energy spectrum of the QP due to the superfluid flow around
vortices. This idea was later developed  to calculate the density of
states, specific heat and thermal conductivity in the mixed state of $d$-wave 
superconductors. Within this
framework \cite{kub1,kub2,hir1} it is possible to explain
 the specific heat \cite{mol} and the thermal conductivity behavior
in a magnetic field perpendicular to the CuO$_2$ planes at low enough temperatures \cite{chi}. 

However, experiments on the field-angle dependence of  the thermal conductivity
with magnetic field applied parallel to the CuO$_2$ planes on YBa$_2$Cu$_3$O$_7$ single crystals (Y123)
 were interpreted assuming Andreev scattering  (AS) of QP by vortices \cite{sal,yu,aub}
taking into account the $d-$wave symmetry of  the order parameter.
We think that this issue needs to be reconsidered experimentally to check whether there is
clear evidence from thermal transport supporting either a DS effect on the energy spectrum of
QP and/or AS. Furthermore, we note that there are no direct measurements 
published on  the transverse thermal 
conductivity $\kappa_{xy}$ for magnetic fields applied parallel to the
CuO$_2$ planes, although the pioneer experiment by Yu et al. \cite{yu}
provides an idea about the possible field-angle dependence of the transverse component.
Nowadays there is consent that $\kappa_{xy}$ is a suitable property to check theoretical expressions of 
electronic thermal transport
 since it is free from  non-electronic contributions.
In this letter we report on the magnetic field $B$, field-angle $\theta$ and temperature $T$ dependence of 
$\kappa_{xy}$ for fields applied parallel to the CuO$_2$ planes. 

The magnetic field, field-angle and temperature dependence of $\kappa_{xy}$ were
studied on two Y123 single crystals. One sample was an optimally doped  twinned 
single crystal with dimensions (length $\times$  width $\times$
thickness) $0.83 \times 0.6 \times 0.045~$mm$^3$ and critical temperature $T_c =93.4~$K \cite{tal,oca1}. 
The second sample was an untwinned Y123 single crystal  with
dimensions $2.02 \times 0.68 \times 0.14~$mm$^3$ and $T_c = 88~$K. 
The heat current $J$ was along the longest axis of the crystal (in case of the
untwinned (twinned) crystal it was parallel to its $a$-axis ($a/b$-axes)). The magnetic field was applied 
parallel to the CuO$_2$ planes within a misalignment angle of $\pm 0.5^0$. 
An in-situ  rotation system enabled us the measurement of 
the  conductivity as a function of the angle $\theta$ defined between 
the applied field and the heat current direction. The used temperature gradient $\nabla_x T \le 300~$K/m
was kept constant along the
heat current direction $x$ during all scans in order
to calculate the transverse component of the thermal conductivity using the
relation
\begin{eqnarray}
\kappa_{xy}(T,B,\theta) = \kappa_{xx}(T,B,\theta) \frac{\nabla_y T}{\nabla_x T} 
\simeq \nonumber \\
\kappa_{xx}(T,B) \frac{\nabla_y T}{\nabla_x T} \propto -\frac{J \nabla_y T}{(\nabla_x T)^2}\,.
\label{kxy}
\end{eqnarray}
In this equation we used the experimental fact that the oscillation amplitude of the longitudinal
component of the thermal conductivity $\kappa_{xx}$ - which has been shown to have
fourfold symmetry at $T/T_c < 0.26$ \cite{aub}- is much smaller than the longitudinal 
temperature gradient $|\nabla_x T|$ \cite{oca2}. 
Therefore $\kappa_{xy}$ shows the same field-angle dependence as $\nabla_y T$.

The longitudinal and transverse temperature gradients 
were measured using a
previously field- and temperature-calibrated type E thermocouples with a dc-picovoltmeter \cite{iny}. 
We obtained similar results in
both crystals for the field-angle and field dependence of $\kappa_{xy}$. 
The temperature dependence of $\kappa^0_{xy}$ as well as its absolute value
depends, however,  on the sample. This result appears to be related to 
 different impurity scattering rates. 
The measured behavior of the longitudinal thermal conductivity as a function of
temperature indicates a stronger impurity scattering for the untwinned crystal probably 
related to its oxygen deficiency.
\begin{figure}
\centerline{\psfig{file=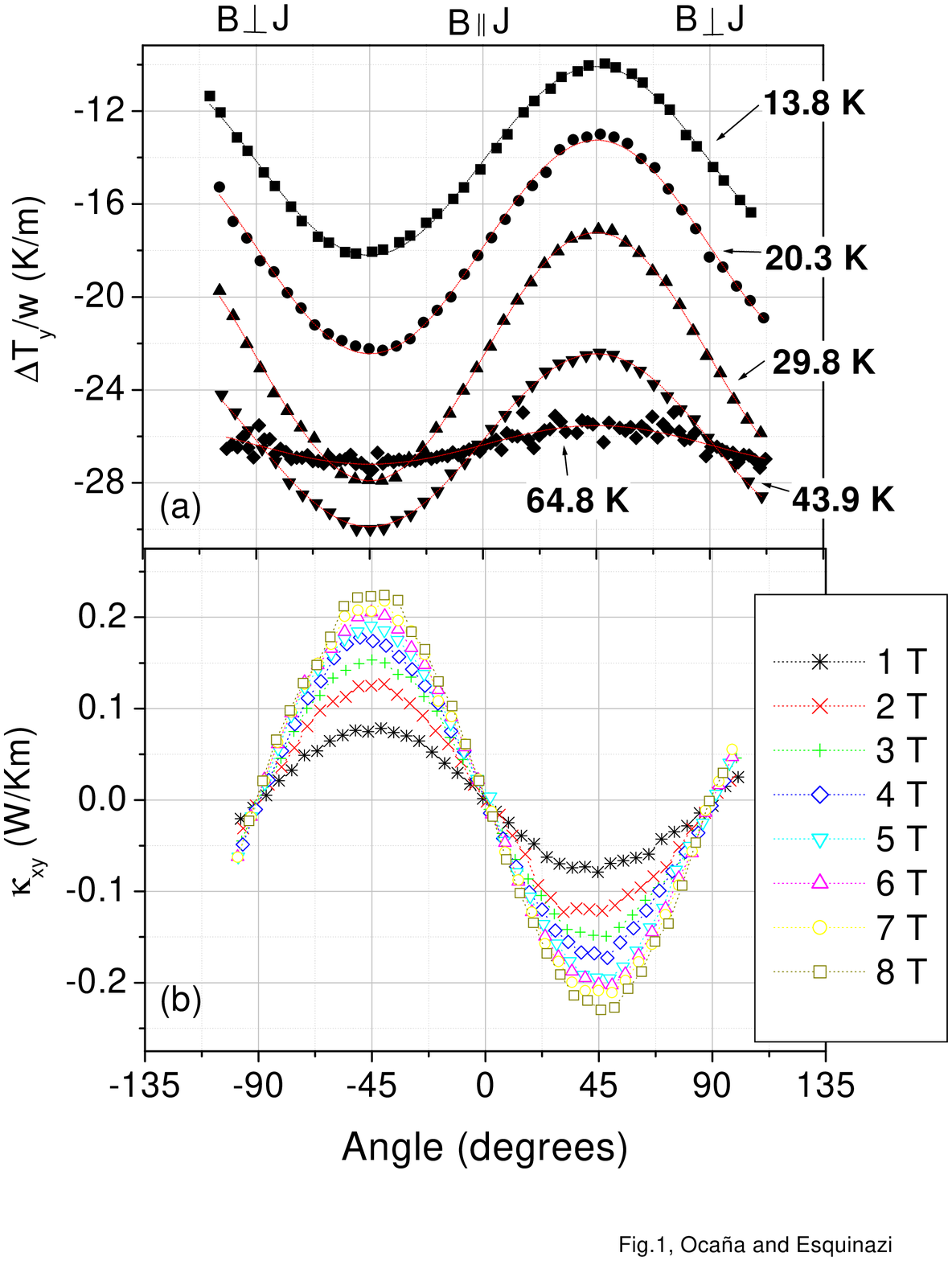,width=85mm}}
\caption{(a) Angle dependence of the transverse temperature gradient at a fixed field
of 8 T and at different temperatures below $T_c$ for the twinned crystal. Continuous lines
are fits to the function $\sin(2\theta)$. (b) Angle
dependence of the transverse  conductivity at $T = 29.8~$K at different 
applied fields parallel to the CuO$_2$ planes for the twinned crystal. If the heat current
direction is along  $+\widehat{x}$ and a {\em positive}  $90^\circ$ angle is along 
$+\widehat{y}$, then there is a positive transverse thermal gradient for $B$ at $+45^\circ$.} 
\label{angle}
\end{figure}

Figure \ref{angle}(a) shows the transverse temperature gradient $\nabla_y T$ as a function
of the angle $\theta$ at different temperatures at a constant field $B = 8~$T.
A clear twofold pattern is measured as the magnetic field is rotated parallel to the CuO$_2$ planes. 
Due to the enhancement of the inelastic scattering rate of QP with increasing temperature
 the angle dependence cannot be resolved above 70~K within the experimental error. The 
background in each curve of Fig.~\ref{angle}(a) is due to the 
experimental misalignment in the position of the sensors of the thermocouple used to measure
$\nabla_y T$ since its
temperature dependence follows roughly that of $\nabla_{x}T$ (the
transverse thermocouple measures partially the longitudinal gradient on the sample as well).
Taking into account the measured longitudinal conductivity $\kappa_{xx}$ at $B=0$ 
we estimate a maximum misalignment of $80~\mu$m along the heat current
direction between the ends of the transverse thermocouple sensors.
Thus, since the length and width of both crystals are larger than the
misalignment of the transverse thermocouple, we can consider
our system as a probe of the true transverse signal, in contrast to Ref.~\cite{yu}. 

A proof of this assumption is that within our experimental
resolution the curves keep their background when the 
magnetic field is changed. This result allows us to subtract it
from the curves and to consider only the oscillation amplitude $\kappa^0_{xy}(T,B)$ 
which contains  the 
physical information of the transverse  response of the system.
The angular dependence of $\kappa_{xy}$ obtained in this way is shown in Fig.~\ref{angle}(b) for
different values of the magnetic field. We think that this is the first time
that such a symmetry is unambiguously established for $\kappa_{xy}$; it follows a field-angle dependence
proportional to $\sin(2\theta)$. Based on the published work we assume that
 the twofold field-angle dependence 
reflects the symmetry of a $d-$wave order parameter. Note
 that the amplitude of the angle dependence on the transverse 
gradient is a relatively larger effect than in the longitudinal case 
(e.g. for the twinned crystal at
$B=8~$T and $T=13.8~$K,  $\nabla_y T = 2.1~$K/m for a longitudinal
gradient of $\nabla_x T = 241~$K/m;  the amplitude of the oscillation of the longitudinal 
component remains of the order of one percent or less of $\nabla_x T$). 
This is the reason for the results obtained in the experiment of Yu et al. \cite{yu} where
 only a twofold pattern was measured, not the fourfold one expected for the field-angle dependence
of the longitudinal thermal conductivity \cite{sal,aub}. A comparison between these two
effects will be reported elsewhere \cite{oca2}. 

The observed angular
dependence of the thermal conductivity in Refs. \cite{sal,yu,aub} was interpreted
in terms of AS of QP by vortices screening currents using the 2D
expression of the Bardeen-Richayzen-Tewordt model and assuming a $d-$wave 
order parameter \cite{bar,yu}:
\begin{equation}
\kappa_{\alpha \beta} = \frac{1}{2\pi^2ck_BT^2\hbar^2}\int^\infty_{p_F}{\rm d}^2p \frac{v_{g\alpha}v_{g\beta}
E_{\bf p}^2}{\Gamma(B,{\bf p},T)} {\rm sech}^2 \left(\frac{E_{\bf p}}{2 k_B T} \right) \,,
\label{brd}
\end{equation}
where, in general, $\Gamma (B,{\bf p},T)$ is the relaxation rate given by the sum  of scattering rates of QP
by impurities $1/\tau_0({\bf p})$, by phonons $1/\tau_P({\bf p},T)$, by QP $1/\tau(B,{\bf p},T)$ 
and AS by vortex currents $1/\tau_v(B,{\bf p},T)$ \cite{yu}. $\alpha$ and $\beta$ denote
the $x$ or $y$ directions on the plane of the sample and $v_{g\alpha}$ is the group velocity
along the $\alpha$ direction. All other symbols have the conventional meaning and are described in Ref.~\cite{yu}
in detail.  With
similar parameters as used in \cite{yu,aub}, e.g.  $B_{c2}^{ab} = 500~$T, Ginzburg-Landau
parameter $\kappa = 100$ and a zero-temperature energy gap $\Delta_0(0) = 20$~meV we obtain 
a $\sin(2 \theta)$ angle dependence for $\kappa_{xy}$ as the experiments show. 
In what follows we compare our experimental results and those from
literature - for magnetic fields applied {\em parallel} to the CuO$_2$ planes - and discuss to which 
extent the available data can help to differentiate between the AS and DS contributions proposed by
the existing models.

\begin{figure}
\centerline{\psfig{file=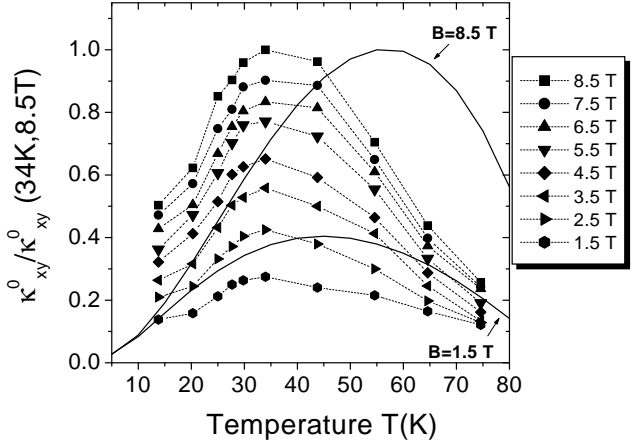,width=85mm}}
\caption{Temperature dependence of the normalized oscillation amplitude 
$\kappa^0_{xy}$ at different applied fields for the untwinned crystal. 
The two continuous curves 
are obtained with (2) with the parameters mentioned in
the text and normalized by $\kappa^0_{xy}(55$K,8.5T).
$\kappa^0_{xy}(34~{\rm K}, 8.5~$T$) = 0.084~$W/Km.}
\label{tdep}
\end{figure}

{\em (1) Angular dependence of $\kappa_{xy}(\theta)$}: We note that the observed angle 
dependence of  $\nabla_{y} T$ agrees with the expected sign if 
AS affects the thermal transport. For example, if $B$ is at $+45^\circ$ the measured $\nabla_y T$,
 see Fig.~\ref{angle}, indicates a higher $T$ at $+90^\circ$, i.e. there is an excess of QP flowing
along the field relative to those moving in the $-45^\circ$ nodal direction.

Recent studies pointed out that a DS
 in the excitation spectrum of QP would produce also a
twofold pattern in the field-angle dependence of  $\kappa_{xy}$
 respectively \cite{won,vek}. In particular, Won and Maki \cite{won,won2} obtained the 
following expression for the transverse conductivity: 
\begin{eqnarray}
\frac{\kappa_{xy}}{\kappa_0} \simeq - \frac{1}{3\pi} \frac{vv'eB}{\Gamma \Delta}
\sqrt{\ln(4\sqrt{\frac{2\Delta}{\pi \hbar \Gamma}})}[ \sin(2\theta) \nonumber \\
 \ln(\frac{2\Delta}{\sqrt{\hbar vv'eB}}) - 0.422 
\sin(2\theta) ] \simeq - \frac{a B \ln(\frac{1}{bB}) \sin(2\theta)}{\kappa_0} \,,
\label{kxywon}
\end{eqnarray}
that applies for $T \ll T_c$,  $B \ll B_{c2}$ and in the clean limit $\tau_0 \gg \hbar/\Delta$,
where $\Delta(\Gamma) \simeq \Delta_0$, this last an impurity independent order parameter. Here 
$\Gamma = \tau_0^{-1}/2$ represents  the impurity
scattering rate, $\kappa_0 = \hbar k_B^2 T n/3\Delta m$,  $v = v_F, v' = v_F \lambda_a/\lambda_c$, $n$ the density 
of QP and $m$ their mass, $b =  \hbar v v' e /4 \Delta^2$. The clean-limit 
expression (\ref{kxywon}) gives a simple field dependence for $\kappa_{xy}^0$ with both, $a$ and $b$, temperature
and sample dependent parameters in which the scattering rate $\Gamma$ of the QP is included.
Within a simple picture the sign of the measured $\nabla_y T$ seems to disagree
 with that expected from the DS:
an excess of QP should be in the nodal direction {\em perpendicular} to the applied field. However, 
the DS affects both the carrier density as well as the scattering rate. This last 
dominates $\kappa_{xy}$ at high enough temperatures \cite{won2}, thus the QP move
easier parallel to the magnetic field and the sign in (\ref{kxywon}) turns out to
be the correct one. The temperature dependence of $\kappa_{xy}$ in the clean limit is still
under development. Although preliminary results \cite{won2} indicate that the main arguments in 
Eq.~(\ref{kxywon}) may remain, we should take (\ref{kxywon}) as a trial function to compare with
our results.

\begin{figure}
\centerline{\psfig{file=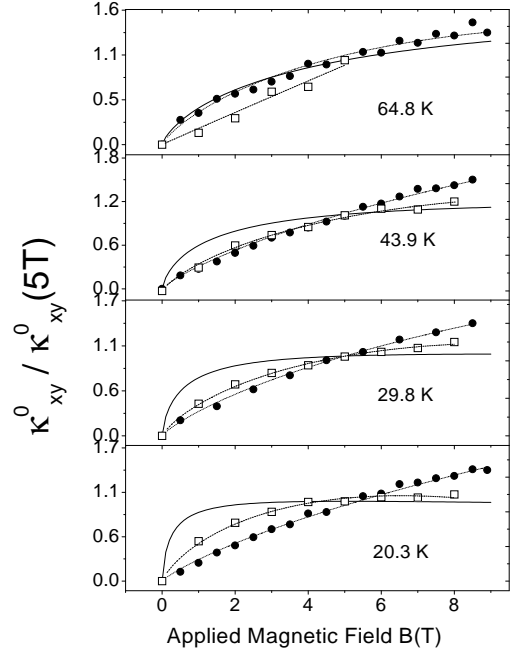,width=85mm}}
\caption{Magnetic field dependence of the normalized oscillation amplitude at different temperatures
for the twinned $(\Box)$ and untwinned $(\bullet)$ crystals. The dashed and dotted
lines are fits of the data to the function $B \ln(1/(bB))$ with  $b$ as a free parameter. 
The continuous line is obtained from Eq.~(\ref{brd}) assuming Andreev and impurity scattering 
with the same parameters
as in Ref.~\protect\cite{yu}.} 
\label{bdep}
\end{figure}

{\em (2) Temperature dependence of $\kappa^0_{xy}$}:
 This dependence  is presented in Fig.~\ref{tdep} for different
 magnetic fields. We note that every point in this
figure represents the measurement of a complete curve as shown in Fig.~\ref{angle} 
and that the experiments below 
20~K have to be performed in a field-cooled procedure to avoid pinning effects
 \cite{aub,tal}. The results show a maximum in  $\kappa^0_{xy}$
at a field-independent temperature of $T \sim 35~$K. $\kappa_{xx}(T)$ shows a 
maximum at a similar temperature, also for 
fields applied parallel to the c-axis of the crystal. These facts  indicate a common origin for the observed maxima: 
a competition takes place between a decreasing  inelastic scattering rate 
due to the opening of the energy gap and the decrease in the number
of QP with decreasing temperature \cite{yu2,hir2}. 
Therefore, to fit the observed $T-$dependence we need the rates $1/\tau_P({\bf p},T)$ and
 $1/\tau(B,{\bf p},T)$ to include in $\Gamma$, see Eq.~(\ref{brd}). 
As seen in Fig.~2, with the above mentioned parameters the AS model
and {\em without}
 $1/\tau_P({\bf p},T)$ and $1/\tau(B,{\bf p},T)$ does not provide a satisfactory fit of the data.
Unfortunately, the exact expressions as well 
as the values of the necessary parameters of the rates are not known with certainty to allow a  
confidential result from the fits. 
Then and since Eq.~(\ref{kxywon}) is strictly valid 
at low $T$, from the observed $T-$dependence
it is not straightforward to conclude which of the proposed mechanisms predominates.
\begin{figure}
\centerline{\psfig{file=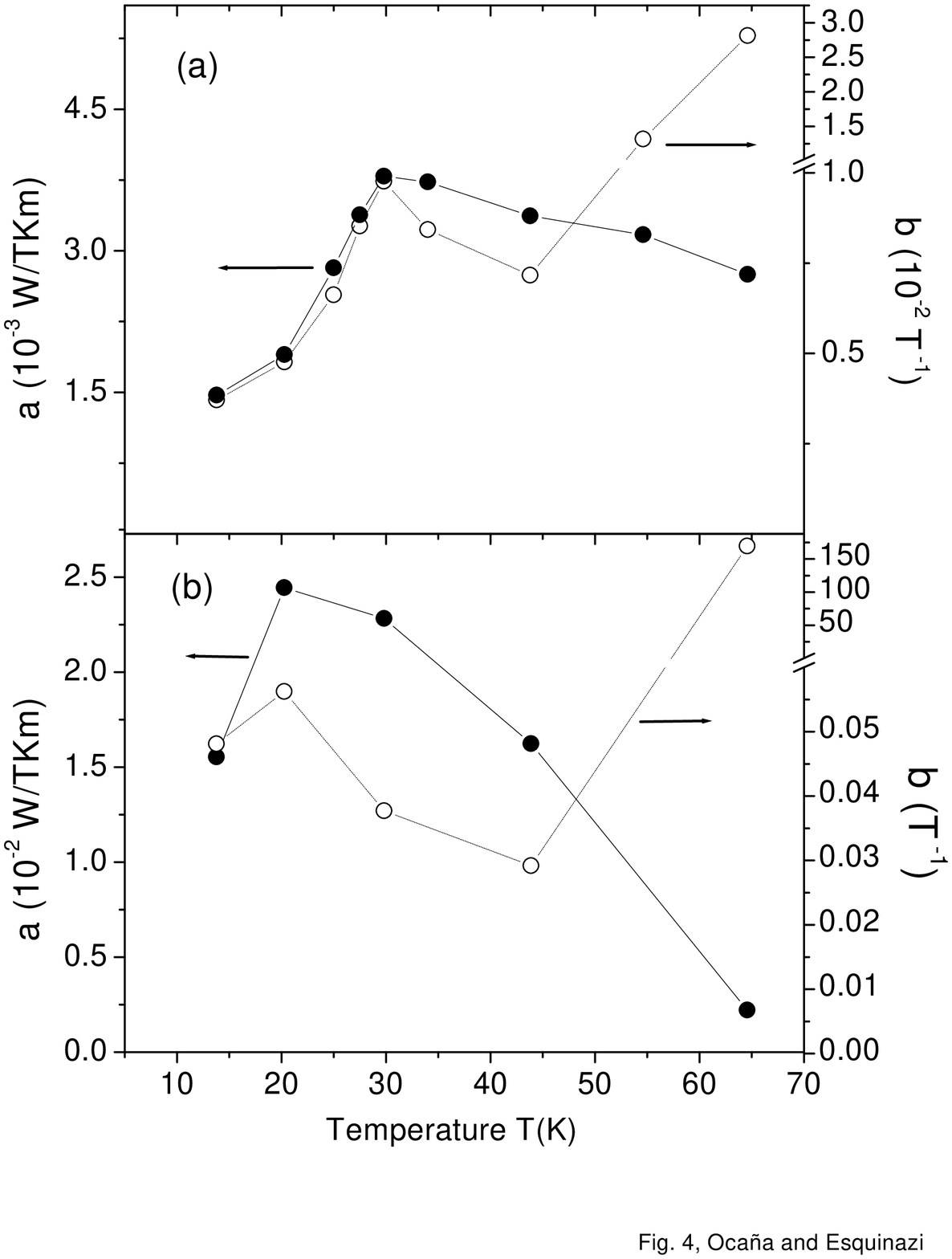,width=80mm}}
\caption{Temperature dependence of the parameters $a (\bullet)$ and $b (\circ)$ obtained from the 
fits to Eq.~(\protect\ref{kxywon}) (see Fig.~\protect\ref{bdep}) for the untwinned (a) and twinned (b) crystals. The 
parameter $a$ is obtained taking the absolute value of $\kappa^0_{xy}(T)$.} 
\label{ab}
\end{figure}

{\em (3) Field dependence of  $\kappa^0_{xy}$}: This dependence  is
presented in Fig.~\ref{bdep} for the two samples at different constant temperatures. 
Equation~(\ref{kxywon}) describes satisfactorily the field 
dependence of $\kappa^0_{xy}$ from the lowest
measured temperature $T \sim 13~$K. The temperature dependence of the fit 
parameters $a$ and $b$ are presented in Fig.~\ref{ab}.
We note that $a$ and $b$ show a similar $T-$dependence for the two samples but different
absolute values.
From the  parameter $a$   obtained at $T \le 20~$K  we calculate the scattering time
$\tau_0 \simeq 0.34 (2.9)~$ps for the untwinned (twinned) crystal assuming 
 $\Delta = \Delta_0(0) = 20$ meV and $v \simeq 5 v' \sim 2 \times 10^5$m/s.
 With these values we obtain a reasonable agreement for the $b$ parameter of the
untwinned crystal. Note, however, that the predicted scattering-rate dependence in $b$
is not only too small in comparison to the factor 10 difference between samples, but is qualitatively 
different, i.e. $b$ increases with $\Gamma$ \cite{approx}.
  Assuming  Andreev and impurity scattering
rates in (\ref{brd})  we cannot reproduce satisfactorily the observed field dependence of 
Fig.~\ref{bdep} in a broad temperature range even varying generously the fitting parameters.
In Fig.~\ref{bdep} we show the results of this calculation using similar
parameters  as indicated above. At high $T$ both
models provide similar good fits to the  data.

{\em  (4) Field and angular dependence of  $\kappa_{xx}$}: In the temperature region of our work 
$\kappa_{xx}$ decreases with field (see for example Fig.~15 in \cite{tal}). 
This effect as well as the angular fourfold pattern \cite{aub}
appear to be accounted for by the AS mechanism 
and at odds with the DS predictions since in  the clean limit (dirty limit in 
\cite{kub1}) a DS  increases $\kappa^{el}_{xx}$ and gives the incorrect sign for the
fourfold pattern \cite{won}. However, two new approaches can provide an explanation
of the results within the DS effect. First, high $T$ and the influence of scattering processes
revert the field dependence of $\kappa^{el}_{xx}$ (as shown already for 
the superclean limit in \cite{kub1}) and the sign of the fourfold pattern \cite{won2}.
Second, we note that experiments measure the total conductivity in which 
phonons also contribute as heat carriers, i.e. $\kappa_{xx}(T,B)= \kappa^{el}_{xx}(T,B) +
\kappa^{ph}_{xx}(T,B)$. In general it is assumed that $\kappa^{ph}_{xx}$ is field independent. 
This is not necessarily true. Note that  none of the separation methods used in literature is accurate enough
to  prove that $\kappa^{ph}_{xx}$ is strictly field independent.  Due to the phonon-electron interaction a 
decrease of  $\kappa^{ph}_{xx}$  
of a few percent, for fields of a few teslas and parallel to the planes, is possible. 
In this case the DS may  {\em decrease} the total conductivity with field due to its influence to
the phonon contribution \cite{won2}.   This contribution would provide also 
the correct sign of the fourfold pattern.

{\em (5) Angular dependence of the specific heat}:  Recent specific heat results indicate the absence of 
a fourfold symmetry in its angular dependence, expected if the DS changes 
the QP density of states \cite{wang}. It is still unclear, however, how large this effect
should be  \cite{won} as well as the influence of extra effects that may not allow an accurate
test of the expected fourfold symmetry in the specific heat \cite{wang}. 

In summary, 
a clear twofold  angle dependence has been obtained in the whole measured temperature 
and field range for the transverse thermal conductivity. Its oscillation 
amplitude $\kappa^0_{xy}$ follows $a(T) B \ln(1/b(T)B)$ with the sample-dependent 
parameters $a$ and $b$. Rigorously speaking none of the available models explain satisfactorily all the details
of the experimental data. The expected influence of the DS casts doubts whether the AS mechanism is the
only and appropiate one to understand $\kappa_{xy}(T,B)$. We think that a further development
 of both models is necessary before a clear answer can be given.

\begin{acknowledgments}
This work is supported by the DFG under Grant DFG Es 86/4-3 and partially supported
by the DAAD. We thank Y. Kopelevich, H. Suderow, A. Goicochea, 
Yu. Pogorelov, M. A. Ramos,  J. G. Rodrigo,  S. Vieira
for fruitful discussions. We are indebted  with K. Maki for enlightening discussions and
for providing us with unpublished theoretical results.
\end{acknowledgments}


\begin{references}
\bibitem{vol} G. E. Volovik, JETP Lett. {\bf 58}, 6 (1993).

\bibitem{kop} N. B. Kopnin, G. E. Volovik, JETP Lett. {\bf 64}, 690 (1996).
\bibitem{kub1} C. K\"ubert and P.J. Hirschfeld, Phys. Rev. Lett. {\bf 80}, 4963 (1998).
\bibitem{kub2} C. K\"ubert and P.J. Hirschfeld, Solid State Commun. {\bf 105}, 459 (1998).
\bibitem{hir1}P. J. Hirschfeld, cond-mat/9809092.
\bibitem{mol}K. Moler et al., Phys. Rev. Lett. {\bf 73}, 2744 (1994).
\bibitem{chi}M. Chiao et al., Phys. Rev. Lett. {\bf 82}, 2943 (1999).
\bibitem{sal}M.B. Salamon, F. Yu and V. N. Kopylov, J. Superconductivity {\bf 8}, 449 (1995).
\bibitem{yu}F. Yu et al., Phys. Rev. Lett. {\bf 74}, 5136 (1995); F. Yu et al.,
     Phys. Rev. Lett. {\bf 75}, 3028 (1995). See also the comment by  R.A. Klemm et al., Phys. Rev. Lett.
    {\bf  77}, 3058 (1996) and the authors reply by  F. Yu et al., Phys. Rev. Lett. {\bf 77}, 3059 (1996).
\bibitem{aub}H. Aubin et al., Phys. Rev. Lett. {\bf 78}, 2624 (1997).
\bibitem{tal}A. Taldenkov, P. Esquinazi, and K. Leicht, J. Low Temp. Phys. {\bf 115}, 15 (1999).
\bibitem{oca1}R. Oca\~{n}a et al., cond-matter/0004067, to be published in J. Low Temp. Phys. (2001).
\bibitem{oca2}R. Oca\~{n}a and P. Esquinazi, to be published.
\bibitem{iny}A. Inyuskin, K. Leicht, and P. Esquinazi, Cryogenics {\bf 38}, 299 (1998).
\bibitem{bar}J. Bardeen, G. Rickayzen and L. Tewordt, Phys. Rev. {\bf 15}, 982 (1959). 
\bibitem{won}H. Won and K. Maki, cond-mat/0004105.
\bibitem{vek}I. Vekhter and P. J. Hirschfeld, cond-mat/9912253.
\bibitem{won2}H. Won and K. Maki, to be published.
\bibitem{yu2}R.C. Yu et al., Phys. Rev. Lett. {\bf 69}, 1431 (1992).
\bibitem{hir2}P.J. Hirschfeld and W.O. Putikka, Phys. Rev. Lett. {\bf 77}, 3909 (1996).
\bibitem{approx} 
This dependence is obtained within a logarithmic approximation \protect\cite{won};  further
development of the theory is necessary to clarify the scattering-rate dependence of $b$.
\bibitem{wang} Wang et al., cond-mat/0009194.

\end{references}
\end{document}